\documentclass[aps,prl,twocolumn,groupedaddress]{revtex4-1}
\usepackage{graphics}
\usepackage{graphicx}
\usepackage{epstopdf}
\usepackage{nicefrac}
\usepackage{xcolor}
\graphicspath{./fig/}

\begin{document}

\title{Novel subbands in the doped two-orbital Kanamori-Hubbard model}

\author{K. Hallberg}
\email[]{karen@cab.cnea.gov.ar}
\author{Y. N\'u\~nez-Fern\'andez}
\affiliation{Centro At{\'o}mico Bariloche and Instituto Balseiro, CNEA and CONICET, 8400 Bariloche, Argentina}

\date{\today}

\begin{abstract}

We calculate and resolve with unprecedented detail the local density of states (DOS) and momentum-dependent spectral functions at zero temperature of one of the key models for strongly correlated electron materials, the degenerate two-orbital Kanamori-Hubbard model, by means of the Dynamical Mean Field Theory which uses the 
Density Matrix Renormalization Group as the impurity solver. When the system is hole doped, and in the presence of a finite inter-orbital Coulomb interaction we find the emergence of a novel holon-doublon in-gap subband which is split by the Hund's coupling. We also observe new interesting features in the DOS like the splitting of the lower Hubbard band into a coherent narrowly dispersing peak around the Fermi energy, and another subband which evolves with the chemical potential. We characterize the main transitions giving rise to each subband by calculating the response functions of specific projected operators and by comparing with the energies in the atomic limit. The detailed results for the spectral functions found in this work pave the way to study with great precision the microscopic quantum behavior in correlated materials.

\end{abstract}


\maketitle

{\it Introduction \textemdash}
Materials with strong electron-electron correlations due to interactions in local orbitals, like transition-metal oxides with partially filled $d$ or $f$ shells, are among the most interesting problems in condensed matter physics. 
Their fascinating properties like high temperature superconductivity, colossal magnetoresistance, correlation-driven metal-insulator transitions, or heavy fermion behavior, and their sensitivity to external fields make them attractive candidates for applications. In order to achieve a microscopic understanding of the physics underlying their complex behavior, it is important to count with numerical tools which should be precise enough to capture the relevant details to be able to compare with experiments. 

Among the most promising theoretical methods to tackle these materials we have the Dynamical Mean Field Theory (DMFT) \cite{review,georges,metzner} which consists on a mapping of lattice models onto quantum impurity models subject to a self-consistency condition. 
Some recent theoretical studies, for example, analyze multiplets in photoemission experiments in chalcogenides and pnictides \cite{yeehaule}, the role of atomic states in interacting intermediate valence systems \cite{multiplets}, the electronic structure of heavy-fermion compounds \cite{haule} and pnictides \cite{hauleshim} and the competition of different interactions like local Coulomb of Hund interactions in multi-orbital models \cite{stadler}. The development of more sophisticated numerical methods as the impurity solver, which is the bottleneck of the DMFT, such as using the Density Matrix Renormalization Group (DMFT+DMRG) \cite{garcia,EPLreview,Frontiers}, 
and later developments \cite{wolf,ganahl,bauernfeind} 
allowed for the calculation of spectral densities with higher precision. Their advantage over other methods like the ones based on Quantum Monte Carlo\cite{gull}.
is that they are able to tackle zero temperature, larger number of bath sites, any energy scale and the convergence in the real frequency axis directly without having to resort to ill-posed analytic continuations from the Matsubara axis.

The DMFT+DMRG, was used to study the half-filled multiorbital Kanamori-Hubbard model, which includes intra- and inter-orbital Coulomb interactions as well as a ferromagnetic Hund coupling between the orbitals.
In particular, for the two-orbital model on the Bethe lattice, well defined quasiparticle peaks were observed in the local density of states (DOS) for the half-filled system in the quasi-localized metallic state close to the Mott transition and in the orbital selective phases (OSP). These peaks were characterized as formed by inter-orbital holon-doublon bound states \cite{Yurielhd}. Subsequent papers confirmed their existence using other methods, like slave particles \cite{Yashar}, extensive numerical calculations for the three orbital model also within the DMFT using the Numerical Renormalization Group (NRG) as the impurity solver \cite{vonDelft}, 
or using exact diagonalization to obtain their splitting with $J$ in the OSP \cite{conJ}. 
Photo-induced non-equilibrium holon-doublon excitations have been also obtained in a one-dimensional two-orbital version of the model studied here \cite{rincon} where they were attributed exclusively to the Hund interaction. 
The structure in the inner edges of the Hubbard bands in the half-filled Hubbard model close to the metal-insulator transition \cite{karski,wolf} has also been characterized as holon-doublon excitations \cite{andreashd}, albeit between nearest neighbor sites. 

However, in spite of these achievements, it is still difficult to obtain precise and detailed theoretical electronic structure results in more general situations to compare with experiments, like angular resolved photoemission, inverse photo-emission experiments or optical conductivity measurements.

In this work we revisit the two-band Kanamori-Hubbard model, one of the key models to study strongly correlated materials. Using the DMFT+DMRG numerical technique we obtain the single-particle density of states for the arbitrarily doped case and for an ample range of parameters. We observe a much richer structure than obtained in previous calculations, including the existence of new in-gap subbands which we characterize by calculating the response functions of specific operators and also by comparing to the atomic limit. In particular we find that one of the subbands is formed mainly by inter-orbital holon-doublon excitations which, for high enough ratios of the inter- to intra-orbital Coulomb repulsions, emerge and separate from the upper Hubbard band (UHB). The doped lower Hubbard band (LHB) also splits into subbands composed mainly by well characterized excitations, separated by minigaps. 

These results show that there exists a much more interesting scenario of excitations than previously thought of in these paradigmatic models for correlated systems.


We study the degenerate two-orbital Kanamori-Hubbard model \cite{KanamoriHubbard,reviewGeorges}:

\begin{equation}
H=\sum_{\left\langle ij\right\rangle \alpha\sigma}t_{\alpha}c_{i\alpha\sigma}^{\dagger}c_{j\alpha\sigma} - (\mu -\epsilon )\sum_{i}n_{i} + \sum_{i}\hat{V}_{i}\mbox{,}\label{eq:Hred}
\end{equation}
where $\left\langle ij\right\rangle $ are nearest-neighbor sites in the Bethe lattice, $\alpha=1,2$ are orbital indices, and $\sigma$ the spin index.
The creation and destruction operators are $c^{\dagger}$ and $c$, respectively and $n_i=\sum_{\alpha\sigma}c_{i\alpha\sigma}^{\dagger}c_{i\alpha\sigma}$ the on-site particle number operator. The nearest-neighbor hoppings are $t_1=t_2=0.5$, which is taken as the unit of energy, implying a non-interacting band width in the Bethe lattice of $W=2$. No inter-orbital hybridization is considered.
Here $\mu$ is the chemical potential where $\mu=0$ leads to half-filled bands. This implies that the site energies must take the values $\epsilon=-U/2 - U_2 + J/2$

The on-site interactions $\hat{V}_{i}$ are:
\begin{equation}
\begin{array}{c}
\hat{V_{i}}=U\sum_{\alpha}n_{i\alpha\uparrow}n_{i\alpha\downarrow}+\sum_{\sigma\sigma'}\left(U_{2}-J\delta_{\sigma\sigma'}\right)n_{i1\sigma}n_{i2\sigma'}-\\
-J\left(c_{i1\uparrow}^{\dagger}c_{i1\downarrow}c_{i2\downarrow}^{\dagger}c_{i2\uparrow}+c_{i1\downarrow}^{\dagger}c_{i1\uparrow}c_{i2\uparrow}^{\dagger}c_{i2\downarrow}\right)\\
-J\left(c_{i1\uparrow}^{\dagger}c_{i1\downarrow}^{\dagger}c_{i2\uparrow}c_{i2\downarrow}+c_{i2\uparrow}^{\dagger}c_{i2\downarrow}^{\dagger}c_{i1\uparrow}c_{i1\downarrow}\right)
\end{array}\label{eq:interaction}
\end{equation}
where $J>0$ is the local exchange Hund's coupling and $U$ ($U_2$) is the intra (inter)-orbital Coulomb repulsion between electrons.
We apply the DMFT\cite{review,georges,metzner} to solve the model using the DMRG\cite{white1, karen1} to obtain the impurity's Green's functions \cite{garcia,EPLreview,Frontiers}
on the real axis with a small imaginary offset $\eta\sim 0.1$ using $L=48$ spinfull orbitals. 

We will consider two situations: a) The more simple case in which the Hund coupling is $J=0$ but keeping the intra and inter-orbital interactions $U$ and $U_2$ finite, and also, b)
finite values of $J$ including the rotationally invariant model corresponding to the physical $t_{2g}$ orbitals which means a finite $J$ and $U_2=U-2J$ \cite{reviewGeorges}.

{\it Results \textemdash} First, we present results for $J=0$ while varying $U_2$ (i.e. not necessarily in the rotational invariant case). In all figures we show the results for one of the orbitals since, by symmetry, both orbitals are equivalent.

In Fig. 1 we show the sequence of local single-particle densities of states (DOS) for fixed on-site $U=3.5$ and $\mu=-1.5$ with varying the inter-orbital interaction $U_2$. For $U_2=0$ we have the two well known Hubbard bands: the UHB and the LHB. For this chemical potential the system is slightly hole doped with a finite DOS at the Fermi energy. When the inter-orbital Coulomb interaction $U_2$ is set on we can see a subband separating and emerging from the UHB towards lower energies, and which evolves into an independent well defined subband located within the Mott-Hubbard gap with increasing $U_2$. As we will show below, we identify this subband as formed mainly by local inter-orbital holon-doublon pairs: a holon-doublon band (HDB). The HDB starts separating from the UHB for finite $U_2$ and it's fully visible within the gap for $U_2 > 1$ that is, for interacting energies of the order of or larger than the UHB half-width. The approximate location of this band in the atomic limit is given by the expressions in Table I and indicated by blue arrows in the figure. The HDB continues to move to lower energies with $U_2$ until it overlaps with the LHB. When $U_2=U=3.5$ and $J=0$ the system has SU(4) rotational symmetry and the holon-doublon (HD) excitations are degenerate with the $|\sigma,\sigma' \rangle$ states (see Table I). 
The HDB emerging in the hole-doped situation is an incoherent subband which has a different character than the narrow HD quasi-particle peak stemming from the coherent metallic DOS in the half-filled case \cite{Yurielhd}.

It is interesting to see that the LHB splits into three subbands with $U_2$. Close to the Fermi energy we find a splitting into two subbands: one corresponding to a band centered at $\omega=0$ (LHB1) and another somewhat broader subband, (LHB2) separated from the first one by a minigap. In addition there is a small feature at larger negative energies (LHB3) which also moves with $U_2$. The UHB is always present at the same energy, which is expected for a fixed value of the chemical potential and $U$, albeit with a smaller weight due to the spectral weight lost towards the HDB states.

In Fig. 2 we plot the DOS for particular values of $U$ and $U_2$ while varying the chemical potential $\mu$. For the half-filled case ($\mu=0$) the system is insulating for these local interaction parameters and we can see the well known lower and upper Hubbard bands symmetrically located around $\omega=0$. As soon as the system is doped with $\mu$ we find an in-gap structure appearing in between these Hubbard bands which we identify as the holon-doublon band mentioned before and which is always present as long as the bands are metallic. The approximate location of this band in the atomic limit is, again, given by the expressions in Table I and are indicated by blue arrows in the figure.  

Here we also observe the splitting of the LHB into three subbands, the low energy ones being LHB1 (around $\omega=0$) and LHB2.  The
LHB2 subband crosses the Fermi energy when decreasing $\mu$ (i.e. doping with holes) and continues moving to higher energies with hole doping (follow the green arrows in the figure).
In this figure we also see the structure at $\omega<0$ at larger negative energies (LHB3), also splitting from the LHB, which arises also due to the existence of an inter-orbital interaction $U_2$ and whose location depends on $\mu$ and on $U_2$ (follow the red arrows).

\begin{figure}
 \includegraphics[scale=0.5]{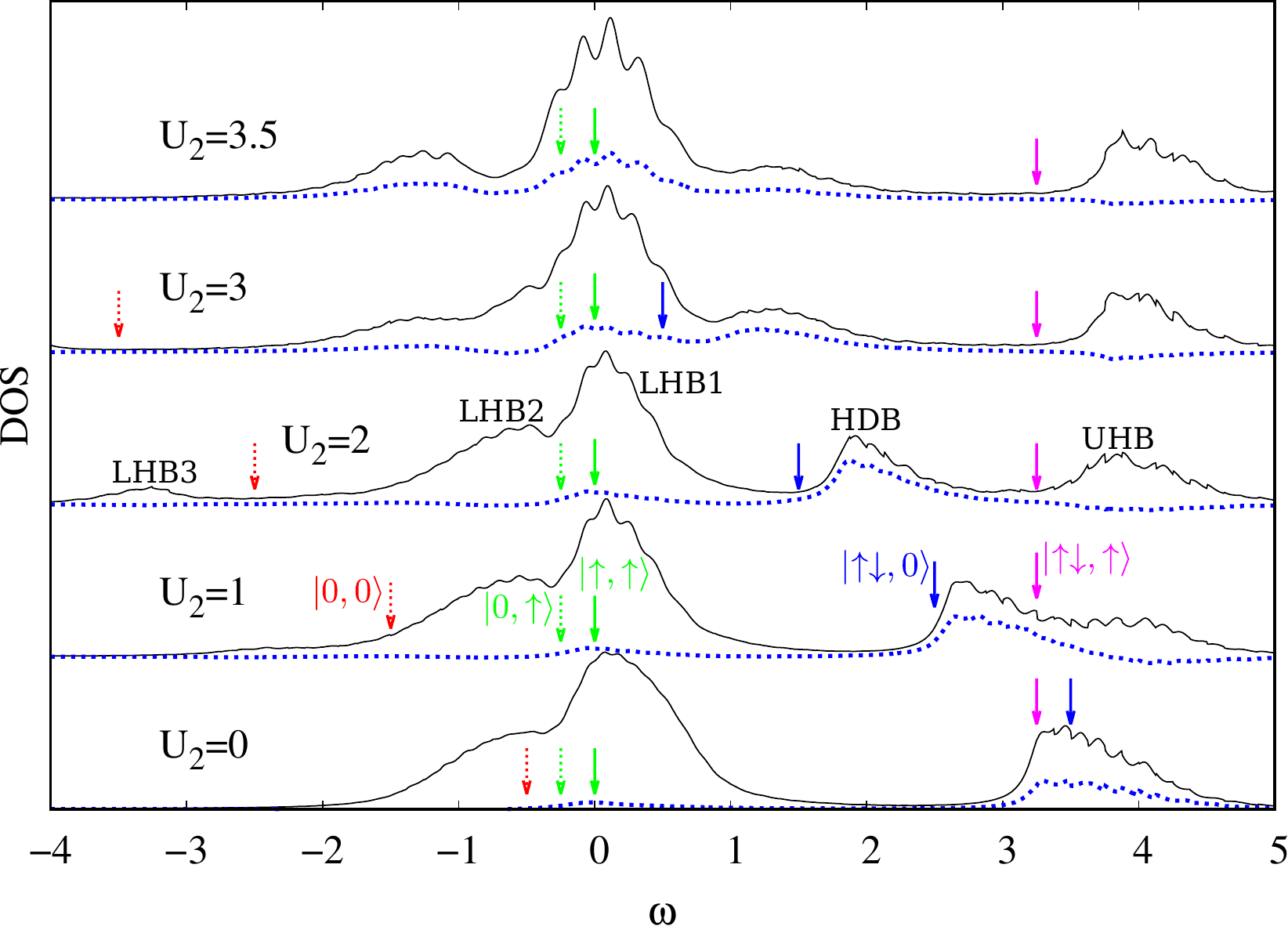}
  \caption{\label{figure1} Single-particle densities of states (DOS) for the two orbital Kanamori-Hubbard model for fixed on-site $U=3.5$, $J=0$, chemical potential $\mu=-1.5$  with varying inter-orbital interaction $U_2$. The energies of each configuration (shown by its representative) in the atomic limit relative to the atomic ground state (arrows at $\omega=0$) are indicated with arrows. The HDB emerging from the UHB with $U_2$ is clearly visible, as well as the splitting of the LHB into three subbands, LHB1, LHB2 and LHB3.  Also shown is $A_{HD}(\omega)$ (blue dotted line).}
 \end{figure}

\begin{figure}
 \includegraphics[scale=0.5]{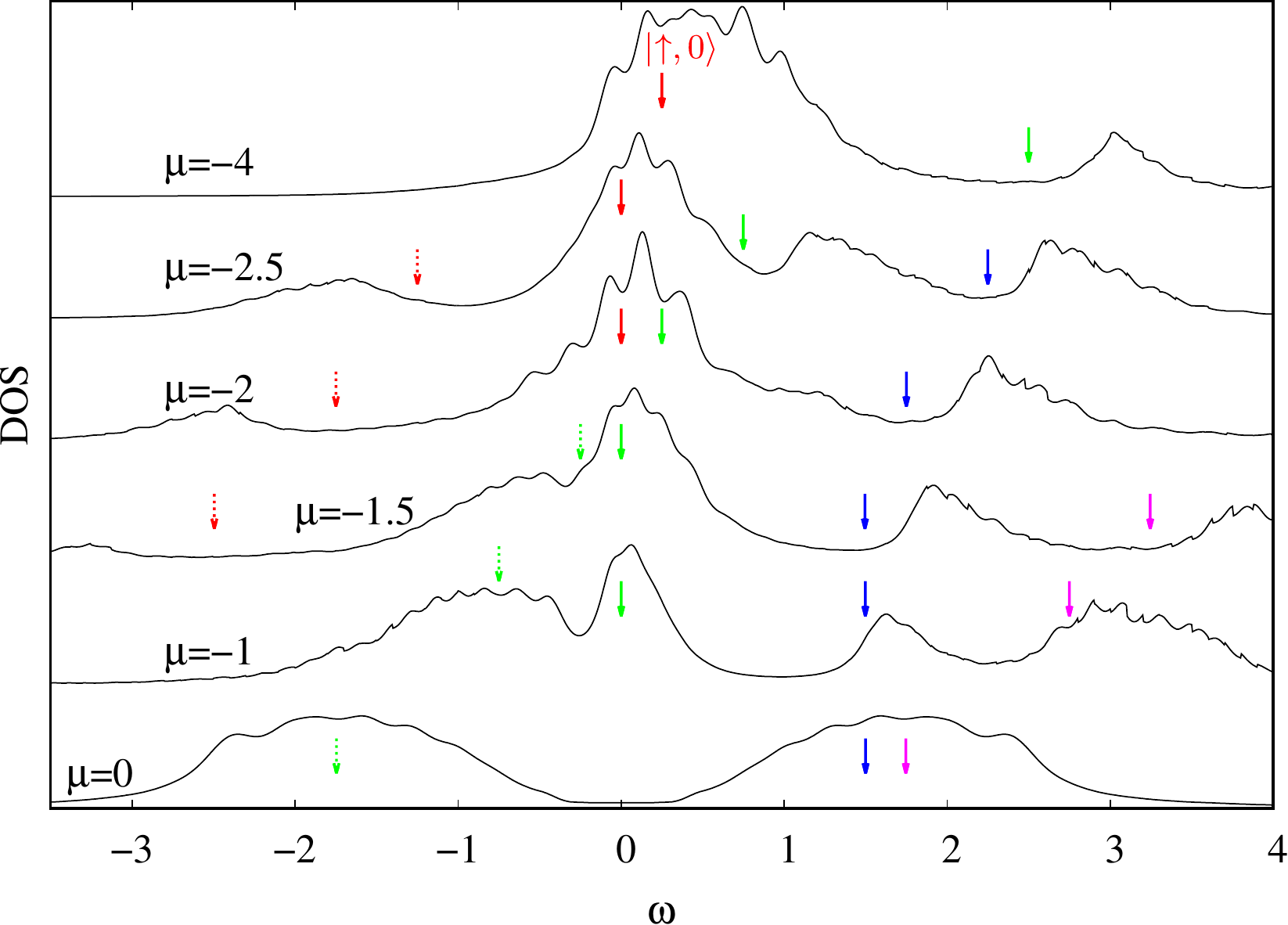}
 \caption{\label{figure2} DOS for fixed on-site $U=3.5$, $J=0$ and inter-orbital $U_2=2$ for several values of the chemical potential $\mu$, from the half-filled insulating case ($\mu=0, \eta=0.2$) to the extremely doped case $\mu=-4$. The arrows (atomic limit energies) have the same color code as in Fig. 1.} 
  \end{figure}

We now consider a finite Hund interaction $J$. In this case we find a much richer structure: the HDB and the LHB2 bands split into two bands separated by $2J$ (see Fig. 3), the former due to the pair hopping term and the latter due to the spin flip term in Eq. 2).
For the rotationally symmetric case in which $U_2=U-2J$, the lower energy states of the HDB coincides with the higher energy ones of the LHB2 (see Table I). 
Splittings due to the Hund's $J$  were also found in Ref. \cite{conJ}, but for the half-filled orbital selective Mott phase (OSMP). We show here that it is not essential to be in an OSMP nor to have a small ratio of hoppings in both bands to observe the HDB or its splittings due to $J$ as stated in that work. 
We also expect that splittings with $J$ should be observed in the model studied in \cite{vonDelft}, however not reported in that work.

\begin{figure}
 \includegraphics[scale=0.5]{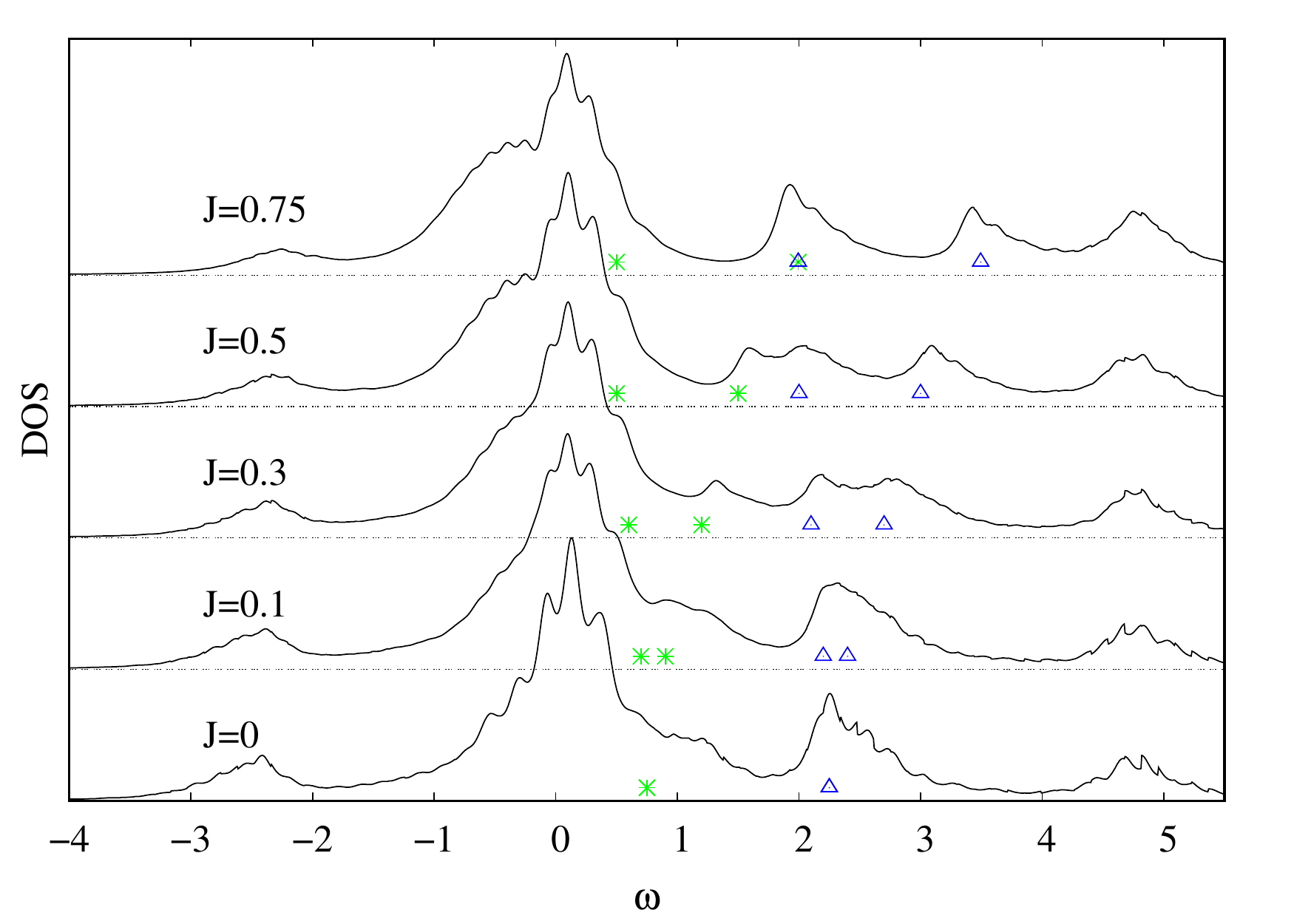}
 \caption{\label{figure3} DOS vs Hund coupling $J$ for fixed on-site $U=3.5$, $U_2=2$ and $\mu=-2$. The SU(4) orbitally symmetric case corresponds to $J=0.75$. The splitting of $2J$ of the HDB and the  $|\uparrow,\downarrow\rangle\pm|\downarrow,\uparrow\rangle$ bands is marked as a guide by blue triangles and by green asteriscs respectively.} 
  \end{figure}

A gross estimation of the energies of these novel excitations can
be done by simple calculations considering the most relevant states
in the two-orbital atomic limit represented as $|s_{1},s_{2}\rangle$
where $s_{1},s_{2}\in\{0,\uparrow,\downarrow,\uparrow\downarrow\}$,
see Table \ref{tab:atomic}. 
The connection of these energies with the DOS
of the lattice is done by considering that the creation or destruction operators,
$c_{1\uparrow}^{\dagger}$ or $c_{1\uparrow}$, applied to the ground state of the lattice at the atomic site 
lead to each one of the atomic configurations of the table. The relative energy
$\epsilon_{|s_{1},s_{2}\rangle}$ of each final configuration $|s_{1},s_{2}\rangle$
with respect to the energy $\epsilon_{0}$ of the atomic ground state
is represented in the DOS at frequency $\omega=\epsilon_{|s_{1},s_{2}\rangle}-\epsilon_{0}$ ($\omega=-\epsilon_{|s_{1},s_{2}\rangle}+\epsilon_{0}$)  for creating (destroying) an electron.
These relative energies are indicated with colored arrows in the figures.
Note that for the half-filled or lightly hole-doped system the atomic ground state is $|\uparrow,\uparrow\rangle$, 
meaning that the values of the Table \ref{tab:atomic} will directly
correspond to excitations of the DOS in this picture. For higher dopings there is a level crossing (c.f. Fig. 2) and the ground state at the atomic level is represented by the state 
$|\uparrow,0\rangle$ so the excitation energies have to be rescaled accordingly.

\begin{table}
\caption{\label{tab:atomic} Relevant representative atomic configurations
and their energy relative to the energy of the atomic state $|\uparrow,\uparrow\rangle$.}

\centering{}%
\begin{tabular}{|c|c|}
\hline 
Representative atomic configuration $|s_{1},s_{2}\rangle$ & $\epsilon_{|s_{1},s_{2}\rangle}$ \tabularnewline
\hline 
\hline 
$|0,0\rangle$ & $U+U_{2}+2\mu$\tabularnewline
\hline 
$|\uparrow,0\rangle$ & $\nicefrac{U}{2}+\nicefrac{J}{2}+\mu$\tabularnewline
\hline 
$|\uparrow,\uparrow\rangle$ & $0$\tabularnewline
\hline 
$|\uparrow,\downarrow\rangle\pm|\downarrow,\uparrow\rangle$ & $J\mp J$\tabularnewline
\hline 
$|0,\uparrow\downarrow\rangle\pm|\downarrow\uparrow,0\rangle$ & $U-U_{2}+J\mp J$\tabularnewline
\hline 
$|\uparrow\downarrow,\uparrow\rangle$ & $\nicefrac{U}{2}+\nicefrac{J}{2}-\mu$\tabularnewline
\hline 
\end{tabular}
\end{table}

To characterize the excitations we calculate the dynamical response
function for creating or destroying an electron on each of the main
components of the ground state according to Table \ref{tab:atomic}.
Generalizing the operators defined in Ref. \cite{andreashd} allows us to calculate directly the local character of the excitations without having to resort to approximations: 
we define the Green's functions $A_{s_{1},s_{2}}(\omega)=A_{s_{1},s_{2}}^{>}(\omega)+A_{s_{1},s_{2}}^{<}(-\omega)$
with:
\begin{equation}
A_{s_{1},s_{2}}^{>}(\omega)=-\frac{1}{\pi}\Im\langle c_{1\uparrow}(\omega+i\eta-H_{imp}+E_{0})^{-1}X_{s_{1},s_{2}}^{\dagger}\rangle\label{eq:Aq>}
\end{equation}
\begin{equation}
A_{s_{1},s_{2}}^{<}(\omega)=-\frac{1}{\pi}\Im\langle c_{1\uparrow}^{\dagger}(\omega+i\eta-H_{imp}+E_{0})^{-1}X_{s_{1},s_{2}}\rangle\label{eq:Aq<}
\end{equation}
where the expectation is taken for the ground state with energy $E_{0}$
of the DMFT Hamiltonian $H_{imp}$. The excitations are $X_{s_{1},s_{2}}^{\dagger}=P_{s_{1},s_{2}}c_{1\uparrow}^{\dagger}$
and their reverse action $X_{s_{1},s_{2}}=c_{1\uparrow}P_{s_{1},s_{2}}$.
The projector $P_{s_{1},s_{2}}=|s_{1},s_{2}\rangle\langle s_{1},s_{2}|$
is used to select the corresponding atomic configuration $|s_{1},s_{2}\rangle$.
Note that adding all possible configurations gives the total DOS since 
$\sum_{s_{1},s_{2}}P_{s_{1},s_{2}}=1$.

We are particularly concerned about the following excitations (and their
reverse actions) for orbital 1:

(i)  HD states ($|\downarrow,0\rangle\to|\uparrow\downarrow,0\rangle$):
The conspicuous HD band was studied using the following holon-doublon
operator: 
\begin{equation}
X_{\uparrow\downarrow,0}^{\dagger}=n_{1\uparrow}n_{1\downarrow}(1-n_{2\uparrow})(1-n_{2\downarrow})c_{1\uparrow}^{\dagger}
\end{equation}
which projects onto the holon-doublon states $|\uparrow\downarrow,0\rangle$ of the single-particle excitation spectrum.

(ii)  ``$\uparrow\uparrow$'' states ($|0,\uparrow\rangle\to|\uparrow,\uparrow\rangle$):
\begin{equation}
X_{\uparrow,\uparrow}^{\dagger}=n_{1\uparrow}n_{2\uparrow}(1-n_{2\downarrow})(1-n_{1\downarrow})c_{1\uparrow}^{\dagger}
\end{equation}
(iii)  ``$\uparrow0$'' states ($|0,0\rangle\to|\uparrow,0\rangle$):
\begin{equation}
X_{\uparrow,0}^{\dagger}=n_{1\uparrow}(1-n_{1\downarrow})(1-n_{2\uparrow})(1-n_{2\downarrow})c_{1\uparrow}^{\dagger}
\end{equation}

As the response functions (\ref{eq:Aq>}) and (\ref{eq:Aq<}) are not diagonal they can lead to small negative values.  

In Fig. 4 we show a breakdown of the main single-particle excitations and a comparison to the total DOS for orbital 1 (those corresponding to orbital 2 are equivalent). We observe that the HDB is formed mainly by holon-doublon states of the form $|\uparrow\downarrow,0\rangle$ (correspondingly, the DOS of orbital 2 will show states of the form 
$|0,\uparrow\downarrow\rangle$). 
We find that the LHB3 is  formed by transitions $|0,0\rangle \leftrightarrow |\uparrow,0\rangle$ given by $A_{\sigma 0}(\omega)$ while the LHB2 is formed by the transitions $|0,\uparrow \rangle \leftrightarrow |\uparrow,\uparrow\rangle$ given by $A_{\sigma \sigma'}(\omega)$ (for all spin configurations). Instead, the DOS around the Fermi energy contains an admixture of both excitations plus a smaller weight of HD states.
These subbands can be also clearly recognized in Figs. 1,2, 3 and 5. 
For small chemical potentials, the LHB2 subband generated by $A_{\sigma \sigma'}$ appears below the Fermi energy (see Fig. 1) and crosses $\omega=0$ with hole doping (see Fig. 2).
In the inset we show the scaling with number of bath sites, where the subband structure is maintained. 

\begin{figure}
 \includegraphics[scale=0.5]{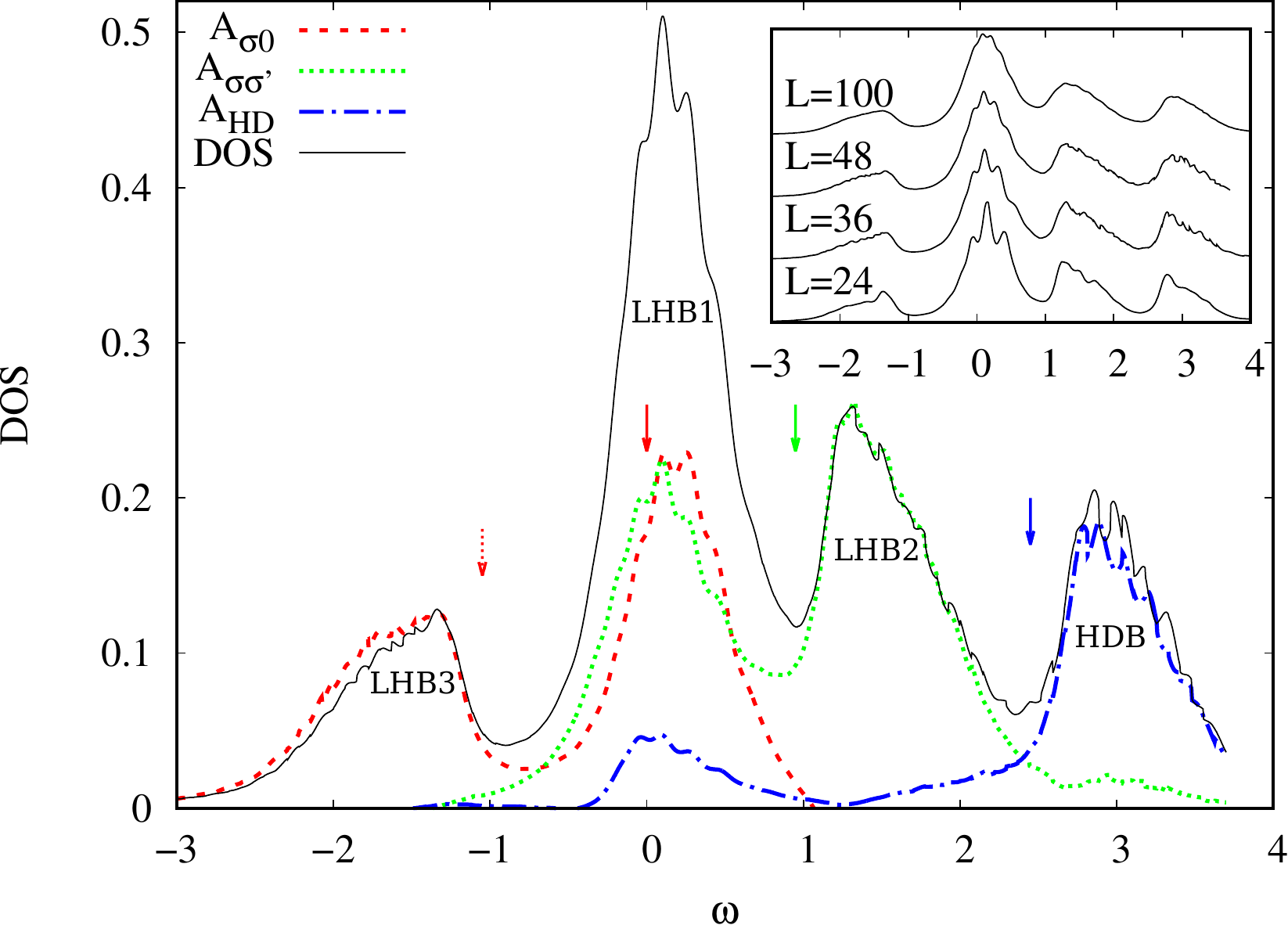}
 \caption{\label{figure4} The decomposition of the DOS for $U=3.5$, $U_2=2$, $J=0$, $\mu=-2.7$ onto the projected excitations $A_{s_{1},s_{2}}(\omega)$ defined in Eqs. (3)-(4). The UHB is out of scale. Clearly resolved are the HDB and the three-fold split LHB subbands. The arrows represent the atomic lmit energies, same color code as in Fig 1. The inset shows the scaling for different bath lengths using $\eta=0.1$.}
  \end{figure}

\begin{figure}
\vspace{0.5cm}
 \includegraphics[scale=0.5]{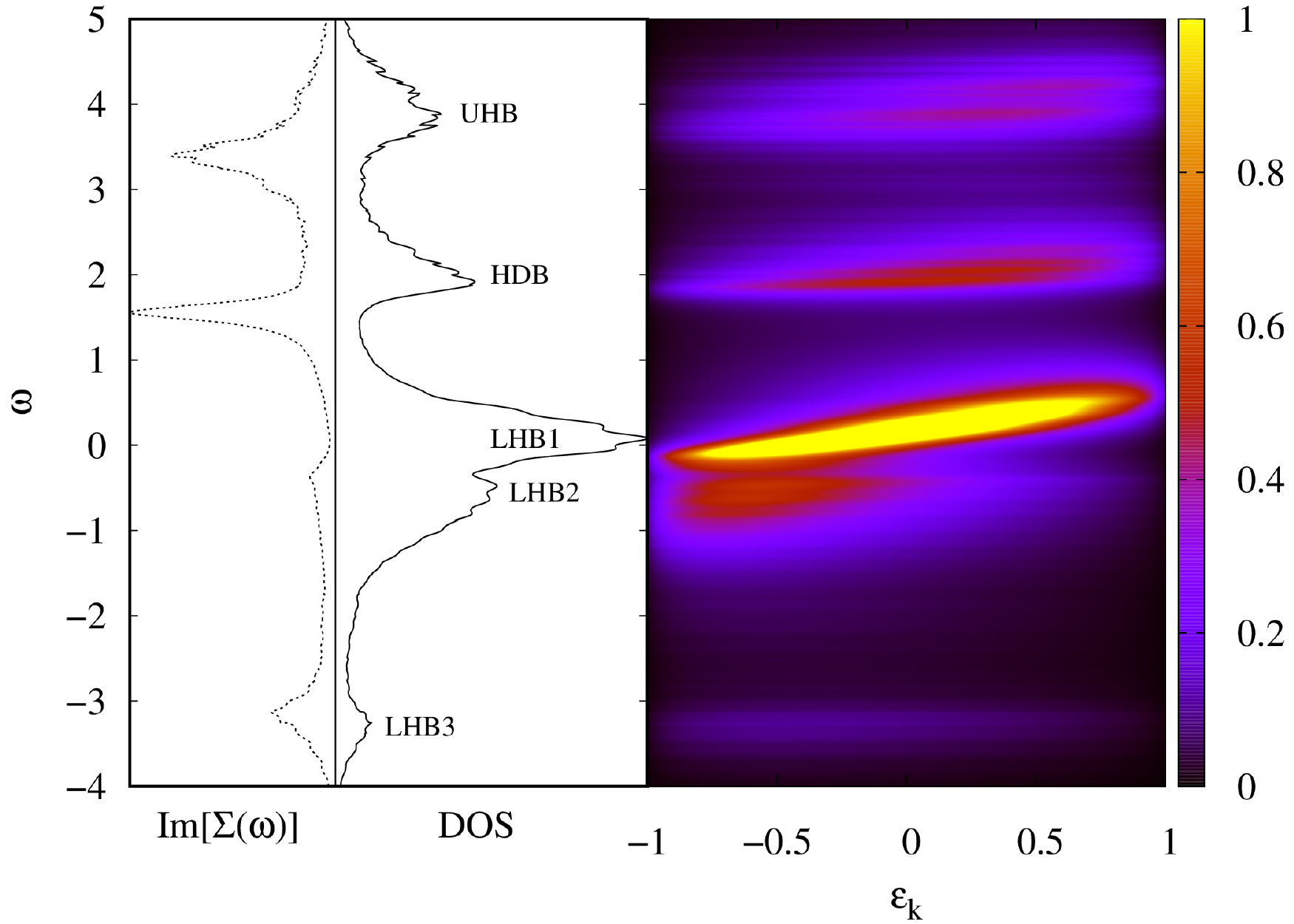}
 \caption{\label{figure5} Spectral function $A(k,\omega)$ for $U=3.5$, $U_2=2$, $J=0$ and $\mu=-1.5$. The band around $\omega=0$ disperses across the Fermi energy. The energy gaps separating each subband are aparent and marked by peaks in the imaginary part of the self energy (also plotted).  } 
  \end{figure}

We have also calculated the momentum-resolved spectral function $A(k,\omega)$ where, within the DMFT the momentum enters via the noninteracting dispersion relation $\epsilon_k$ \cite{review} (see Fig. 5). 
We find that the lowest-lying states around zero (LHB1) disperse across the Fermi energy and are separated by a minigap from the LHB2 (in this case at $\omega<0$) which has a large weight for negative momenta. Also seen are the LHB3 subband at $\omega \sim -3.5$ and the UHB at $\omega \sim 4$.

{\it Conclusions \textemdash}
In this work we report on the existence of novel excitations in a key model to study multiband correlated materials. We were able to observe these excitations, which have passed unnoticed in numerous previous theoretical studies of this model, thanks to the high precision achieved by the numerical method used in this paper (DMFT with DMRG as the impurity solver).

We observe clear subbands within the Mott gap formed mainly by holon-doublon pairs which are pulled down from the upper Hubbard band to lower energies by the inter-orbital Coulomb interaction $U_2$ and are split by the magnetic Hund's interactions by $2J$. The lower Hubbard band also splits into three subbands with $U_2$. One has low weight, corresponding to the completely empty state in the atomic limit, and evolves to negative energies with $U_2$. The other two subbands lie close to the Fermi energy and are separated by a minigap which depends on the doping. One of them is the metallic peak around $\omega=0$. The other one involves states of the form $|s1,s2\rangle=|\sigma,\sigma'\rangle$ in the atomic limit, whose energy increases due to $U_2$. We have also characterized each subband by a careful comparison with the states in the atomic limit and by calculating projected response functions.

The main qualitative results presented here do not depend on particular choices of parameters. We expect these subbands to be robust with a small inter-orbital hybridization, also for the case of three or more orbitals and for other types of lattices or even in one dimension. We also expect similar features to exist for other dopings and related models.

We hope that the results presented here together with the possibility of calculating more precise spectral functions for models of correlated materials will stimulate a closer study of the details of experimental results and, hence, contribute to unveil the complex and elusive microscopic behavior of strongly correlated materials.

\begin{acknowledgments}
This work was completely done in quarantine during the COVID-19 pandemic.
We acknowledge support from projects PICT 2016-0402 and 2018-01546 (ANPCyT) and PIP2015 - 11220150100538 CO (CONICET). We thank G. Kotliar and A. Aligia for useful discussions. 
\end{acknowledgments}


\end{document}